\documentclass[aps,twocolumn,superscriptaddress,groupedaddress]{revtex4}
\usepackage{amssymb}
\usepackage{bm}
\usepackage{amsmath}
\usepackage{graphicx}
\usepackage{epstopdf}
\usepackage{subfigure}
\usepackage{natbib}
\usepackage{epsfig}
\usepackage{amsfonts}
\usepackage{mathrsfs}
\usepackage[toc,page,title,titletoc,header]{appendix}
\usepackage[colorlinks,linkcolor=blue,citecolor=blue,anchorcolor=blue]{hyperref}
\usepackage{dsfont,amsthm,amsbsy}
\usepackage{bbold}
\usepackage{fancyhdr}
\newcommand*\chem[1]{\ensuremath{\mathrm{#1}}} 
\usepackage{chemformula}
\usepackage{floatrow} 
\usepackage{graphicx} 
\usepackage{physics} 
\usepackage{upgreek} 
\usepackage{array}
\newcolumntype{L}[1]{>{\raggedright\let\newline\\\arraybackslash\hspace{0pt}}m{#1}}
\newcolumntype{C}[1]{>{\centering\let\newline\\\arraybackslash\hspace{0pt}}m{#1}}
\newcolumntype{R}[1]{>{\raggedleft\let\newline\\\arraybackslash\hspace{0pt}}m{#1}}

\newcommand*{\vrec}{{\ooalign{\lower.3ex\hbox{$\sqcup$}\cr\raise.4ex\hbox{$\sqcap$}}}}

\begin{document}

\title{The emergence of antiferromagnetic correlations and Kondo-like features\\
in a two-band model for infinite-layer nickelates}
\author{Fangze Liu}
\affiliation{Department of Physics, Stanford University, Stanford, California 94305, USA}
\affiliation{Stanford Institute for Materials and Energy Sciences, SLAC National Accelerator Laboratory, 2575 Sand Hill Road, Menlo Park, CA 94025, USA}

\author{Cheng Peng}
\affiliation{Stanford Institute for Materials and Energy Sciences, SLAC National Accelerator Laboratory, 2575 Sand Hill Road, Menlo Park, CA 94025, USA}

\author{Edwin W. Huang}
\affiliation{Department of Physics and Institute of Condensed Matter Theory, University of Illinois at Urbana-Champaign, Urbana, Illinois 61801, USA}
\affiliation{Department of Physics and Astronomy, University of Notre Dame, Notre Dame, IN 46556, United States}
\affiliation{Stavropoulos Center for Complex Quantum Matter, University of Notre Dame, Notre Dame, IN 46556, United States}

\author{Brian Moritz}
\affiliation{Stanford Institute for Materials and Energy Sciences, SLAC National Accelerator Laboratory, 2575 Sand Hill Road, Menlo Park, CA 94025, USA}

\author{Chunjing Jia}
\affiliation{Stanford Institute for Materials and Energy Sciences, SLAC National Accelerator Laboratory, 2575 Sand Hill Road, Menlo Park, CA 94025, USA}

\author{Thomas P. Devereaux}
\affiliation{Stanford Institute for Materials and Energy Sciences, SLAC National Accelerator Laboratory, 2575 Sand Hill Road, Menlo Park, CA 94025, USA}
\affiliation{Department of Materials Science and Engineering, Stanford University, CA 94305, USA}

\begin{abstract}
We report a determinant quantum Monte Carlo study of a two-band model, inspired by infinite-layer nickelates, focusing on the influence of interlayer hybridization between $3d_{x^2-y^2}$ orbitals derived from $\chem{Ni}$ (or $\chem{Ni}$ and $\chem{O}$) in one layer and rare-earth ($R$) $5d$ orbitals in the other layer, hereafter the $\chem{Ni}$ and $R$ layers, respectively. For a filling with one electron shared between the two layers on average, interlayer hybridization leads to ``self-doped" holes in the $\chem{Ni}$ layer and the absence of antiferromagnetic ordering, but rather the appearance of spin-density and charge-density stripe-like states. As the interlayer hybridization increases, both the $\chem{Ni}$ and $R$ layers develop antiferromagnetic correlations, even though either layer individually remains away from half-filling. For hybridization within an intermediate range, roughly comparable to the intralayer nearest-neighbor hopping $t_{\chem{Ni}}$, the model develops signatures of Kondo-like physics.
\end{abstract}

\maketitle

\section{Introduction}

The recent discovery of superconducting infinite-layer nickelates~\cite{li2019superconductivity} has sparked an inquiry into the relationship between the fundamental physics in infinite-layer nickelates and in cuprates concerning the superconducting mechanism. Various microscopic models have been proposed to capture essential features of infinite-layer nickelates, and a consensus has focused on the role of multi-orbital physics~\cite{Lee2004,Jiang2020,Sakakibara2020prl,Lechermann2020,WuX2020,Nomura2020,hepting2020electronic,Wang2020prb,Werner2020prb,Bhattacharyya2020prb,zhang2020self,zhang2020type,Botana2020PRX,kitatani2020nickelate,kang2023infinite,Liu2021prb,Bjornson2021PRB,Zhang2020prr}. However, controversy still exists over which orbitals matter in a minimum model. 
Given the similar crystal and electronic structures of cuprates and infinite-layer nickelates, it is natural to adapt the effective low-energy model of curprates to describe nickelates, incorporating modifications to account for observed experimental discrepancies between nickelates and cuprates. 
Several studies have explored the superconducting mechanism in doped nickelates~\cite{Jiang2020,zhang2020type}. Unlike in cuprates, where doped holes enters into oxygen sites facilitating Zhang-Rice singlet formation~\cite{zhang1988effective}, in nickelates, it is proposed that doped holes enter into the Ni $3d$ orbitals, which leads to the exploration of a coupling between the localized moment in $3d_{x^2-y^2}$ orbital and the doped hole in $3d_{z^2}$ orbital, with the emergence of Kondo resonance under mean-field calculations~\cite{zhang2020type}. 
The observed upturn behavior of resistivity in the undoped parent compounds and lightly doped nickelates, along with the logarithmic temperature dependence of both resistivity and Hall coefficient at the intermediate temperatures~\cite{li2019superconductivity,ikeda2016direct, osada2020phase, li2020superconducting}, provides clear evidence of magnetic Kondo scattering, which supports the inclusion of Kondo coupling to cuprate-like models~\cite{zhang2020self}. 
Based on the absence of antiferromagnetic long-range order in the parent compounds, attention has also been given to the self-doping effect brought by rare-earth orbitals, which emerges proposals of many cuprate-like models featuring the coupling between $R$ $5d$ orbital and $\chem{Ni}$ $3d_{x^2-y^2}$ orbital ~\cite{Botana2020PRX, Lechermann2020, zhang2020self, hepting2020electronic, been2021electronic}, or $\chem{Ni}$ $3d_{x^2}$ orbital~\cite{hansmann2009turning, held2022phase, PhysRevB.102.100501}.

When focusing on the undoped parent compounds or lightly-doped nickelates, the minimal two-band model, with one $3d_{x^2-y^2}$ band representing physics within the $\chem{NiO_2}$ plane and the other featuring itinerant electrons from the rare-earth elements, has shown to be a strong candidate. Previous density functional theory (DFT)~\cite{Botana2020PRX,hepting2020electronic,been2021electronic} and density-matrix renormalization group (DMRG)~\cite{peng2023charge} studies on this minimal two-band model encourage the view that nickelates are distinct from the cuprates in terms of the specific electronic structure and potential charge density wave (CDW) behavior. The distinction is exemplified by the presence of carriers in both the $\chem{Ni}$ $3d_{x^2-y^2}$ and rare-earth $5d$ layers at a filling of one electron shared between the $\chem{Ni}$ and $R$ layers per site, and when moderately doped, as well as the emergence of charge density waves with a periodicity locked between the $\chem{NiO_2}$ and rare-earth layers~\cite{peng2023charge}. These numerical results have a close correspondence to observations from various spectroscopic experiments~\cite{rossi2021orbital, Goodgee2021pnas, rossi2022broken,tam2022charge,krieger2022charge}. 

In this series of studies, the key to explaining the differences between nickelates and cuprates lies in the interlayer hybridization between the $\chem{Ni}$ layer and the $R$ layer: it leads to ``self-doping", with holes entering into the $\chem{Ni}$ layer and electrons into the $R$ layer, which may partially explain why no antiferromagnetic order has been observed experimentally in the parent compounds. 
However, the small hybridization between $\chem{Ni}$ $3d_{x^2-y^2}$ and $R$ $5d$ orbitals, as estimated by DFT calculations~\cite{been2021electronic}, does not account for the Kondo effect observed in experiments~\cite{li2019superconductivity} within this two-band model. Also, some DFT calculations indicate that the self-doping effect and hybridization between $\chem{Ni}$ $3d$ and $R$ $5d$ bands are reduced by increasing the doping level~\cite{Botana2020PRX, kitatani2020nickelate, Liu2021prb}. 
Nevertheless, studies~\cite{gu2020substantial, foyevtsova2022electride} suggest that the hybridization between Ni $3d$ and interstitial $s$ orbitals can significantly enhance the hybridization between the $\chem{NiO_2}$ plane and itinerant electrons. Furthermore, this hybridization could be further enhanced by rare-earth substitution and/or pressure effects~\cite{been2021electronic, wang2022pressure, guo2021hidden}. In light of this complex scenario, conducting model studies on this two-band model under various levels of hybridization could be insightful.

In this paper, we focus on the influence of interlayer hybridization on the two-band Hubbard model, with self-doping of $1/8$ holes in the $\chem{Ni}$ layer and $1/8$ electrons in the $R$ layer. 
For small interlayer hybridization, this self-doping leads to the absence of long-range antiferromagnetic correlations, but rather the emergence of spin and charge stripe-like features, phenomena supported by our numerical measurements. 
For larger interlayer hybridization, the role of the $R$ layer may go beyond that of just an electron reservoir, potentially leading to the involvement of Kondo-like physics~\cite{hu2017effects,zhang2020self,nomura2022superconductivity}. A shift in the predominantly $R$ band, eventually crossing the Fermi level, leaves the system in an antiferromagnetic insulating state. Antiferromagnetic correlations start to develop in both layers when the interlayer hybridization becomes comparable to the intralayer nearest-neighbor hopping $t_{\chem{Ni}}$, resulting in a Kondo-like coupling close to the strength of intralayer Heisenberg coupling in the $\chem{Ni}$ layer, as predicted by perturbation theory. A notable suppression in the antiferromagnetic spin susceptibility, as well as in the uniform $d$-wave pairing susceptibility mediated by spin fluctuations, has been observed near the cross-over from spin stripes to the antiferromagnetic state. Such non-monotonic behaviors are themselves suggestive of underlying Kondo-like physics.

\vspace{-0.2cm}
\section{Model and Methodology}

We consider a simplified bilayer two-orbital Hubbard model, which resembles a low-energy effective model obtained from Wannier downfolding the DFT-derived band structure for $R$NiO$_2$ ~\cite{hepting2020electronic,been2021electronic}. The Hamiltonian can be written as
\begin{equation}\label{eq:Hamiltonian}
\begin{aligned}
\hat{H} 
& = - \sum\limits_{\ell \in \{\chem{Ni},R\}} t_{\ell} \sum\limits_{\langle ij \rangle, \sigma}  (\hat{c}^{[\ell] \dagger}_{i, \sigma} \hat{c}^{[\ell]}_{j, \sigma} + \hat{c}^{[\ell] \dagger}_{j, \sigma} \hat{c}^{[\ell]}_{i, \sigma}) \\
& \quad - t'_{\chem{Ni}} \sum\limits_{\langle\langle ij \rangle\rangle, \sigma} (\hat{c}^{[\chem{Ni}] \dagger}_{i, \sigma} \hat{c}^{[\chem{Ni}]}_{j, \sigma} + \hat{c}^{[\chem{Ni}] \dagger}_{j, \sigma} \hat{c}^{[\chem{Ni}]}_{i, \sigma})\\
& \quad - t_{\chem{Ni}-R} \sum\limits_{i \sigma} (\hat{c}^{[\chem{Ni}] \dagger}_{i, \sigma} \hat{c}^{[R]}_{i, \sigma} +\hat{c}^{[R] \dagger}_{i, \sigma} \hat{c}^{[\chem{Ni}]}_{i, \sigma}) \\
& \quad + \sum\limits_{i} U_{\chem{Ni}} (\hat{n}^{[\chem{Ni}]}_{i, \uparrow} - \frac{1}{2})(\hat{n}^{[\chem{Ni}]}_{i, \downarrow} - \frac{1}{2}) \\
& \quad - \sum\limits_{\ell \in \{\chem{Ni},R\}} \sum\limits_{i \sigma} \mu_{\ell} \hat{n}^{[\ell]}_{i, \sigma}.
\end{aligned}
\end{equation}
Here, $\hat{c}^{[\ell] \dagger}_{i, \sigma}$ ($\hat{c}^{[\ell]}_{i, \sigma}$) denotes the creation (annihilation) operator of an electron with spin $\sigma$ ($\sigma =\uparrow$, $\downarrow$) at site $i$ within the $\ell$ ($\ell=\chem{Ni}, R$) layer, and $\hat{n}^{[\ell]}_{i, \sigma} = \hat{c}^{[\ell] \dagger}_{i, \sigma} \hat{c}^{[\ell]}_{i, \sigma}$ is the electron density. The overall electron concentration is defined as $n=\sum_{\ell}n_{\ell}$, where $n_{\ell}=\langle \tfrac{1}{N} \sum_{i,\sigma} \hat{n}_{i,\sigma}^{[\ell]}\rangle$ and $N$ is the number of sites in a single layer. $t_{\ell}$ and $t'_{\ell}$ denote the hopping integrals between nearest-neighbor sites and next-nearest-neighbor sites in the $\ell$ layer, respectively, and $t_{\chem{Ni}-R}$ is the hopping integral between the two layers. The parameters $U_{\chem{Ni}}$ and $\mu_{\ell}$ are described in the following paragraph.

We perform numerically exact determinant quantum Monte Carlo (DQMC) simulations~\cite{white1989numerical} on $N=16\times 4$ rectangular clusters (and 2 layers) with periodic boundary conditions along the $x$ and $y$ directions. In this paper, we measure all energies in the unit of $t_{\chem{Ni}}$ and most other parameters, such as $t_{R}=0.5$ and $t'_{\chem{Ni}}=-0.25$, are extracted from Wannier downfolding of the $R$NiO$_2$ bandstructure as in Refs.~\cite{hepting2020electronic,been2021electronic}. We consider the local (on-site) Coulomb interaction in the $\chem{Ni}$ layer to be $U_{\chem{Ni}}=6$. $\mu_{\ell}$ is the chemical potential that controls the electron densities and the on-site energy difference between the $R$ and $\chem{Ni}$ layers, $\epsilon_{\chem{Ni}-R} = \mu_{\chem{Ni}}-\mu_{R}+U_{\chem{Ni}}/2$. While in the undoped parent compound $\chem{NdNiO_2}$, the electron concentration in the $R$ layer is approximately $8\%$ as determined by Hall coefficient measurements~\cite{li2020superconducting}, we choose a commonly selected hole concentration of $12.5\%$ in the $\chem{Ni}$ layer and $12.5\%$ electrons concentration in the $R$ layer to facilitate comparison with measurements in single-band Hubbard model~\cite{huang2018stripe, huang2023fluctuating}. 

In this study, we aim to explore the dependence of various physical properties on the interlayer hopping $t_{\chem{Ni}-R}$, considering a range of values from $0.1t_{\chem{Ni}}$ (typical $\chem{Ni}$ $3d_{x^2-y^2}$ and $R$ $5d$ hybridization from DFT calculations~\cite{hepting2020electronic,been2021electronic}), to $2t_{\chem{Ni}}$ (considering $\chem{Ni}$ $3d$ and interstitial $s$ hybridization~\cite{gu2020substantial}) and even higher values. Up to 80 independently seeded Markov chains and 250000 measurement sweeps are used for each parameter set. The smallest average sign encountered in our simulations is 0.117 for inverse temperature $\beta=t_{\chem{Ni}}/k_{\text{B}} T=6$ and $t_{\chem{Ni}-R}\sim 0.8$. Error bars displayed in DQMC results represent a range that extends one standard error above and below the mean values, as determined by jackknife resampling.

\begin{figure*}[htb!]
\includegraphics[width=0.9\linewidth]{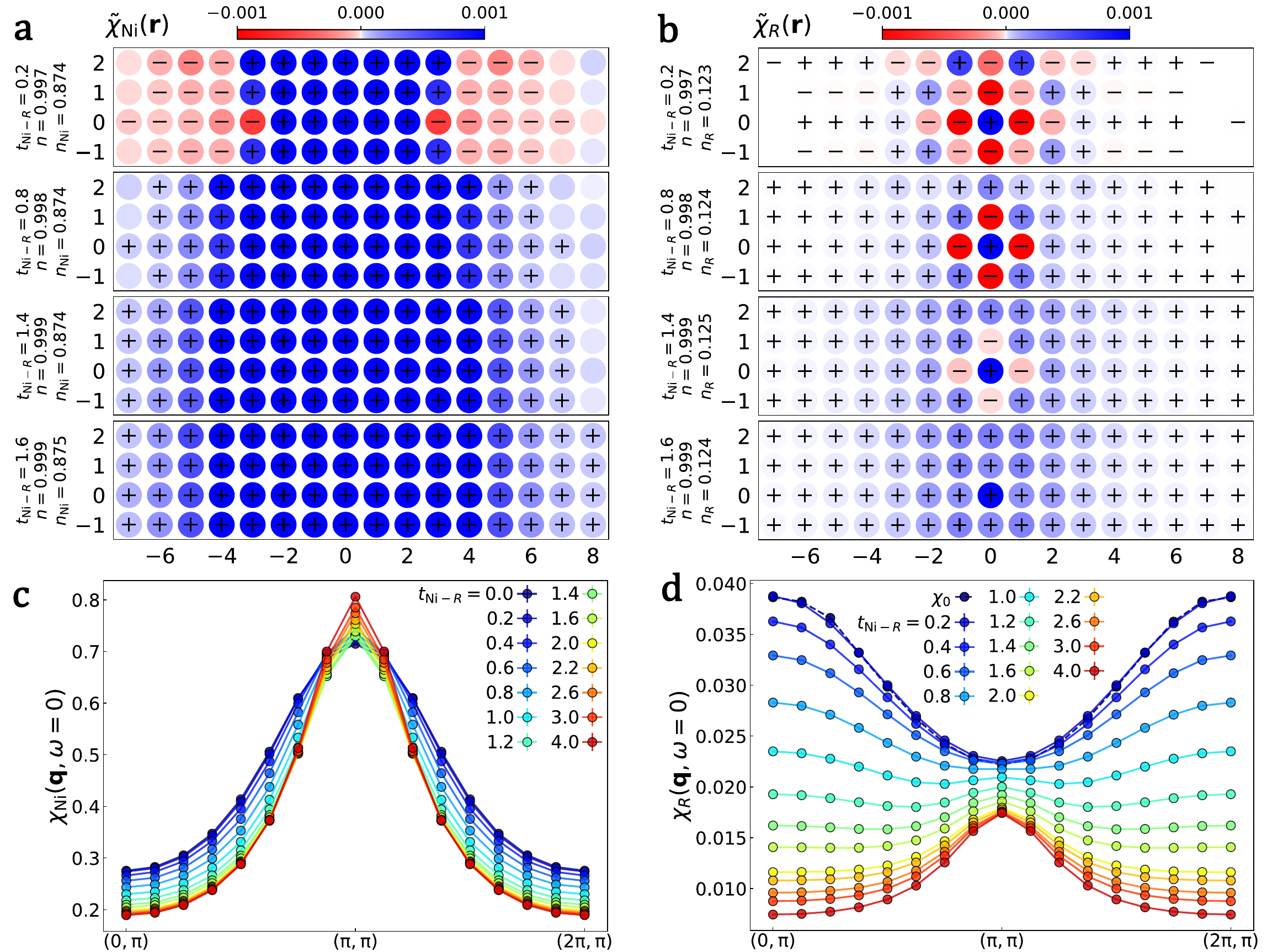}
\caption{\label{fig1} \textbf{Spin susceptibilities in the $\chem{Ni}$ and $R$ layers.} Real-space staggered spin susceptibility for the (a) $\chem{Ni}$ layer and (b) $R$ layer for various $t_{\chem{Ni}-R}$. A $+$ or $-$ is shown when the susceptibilities are nonzero by at least two standard errors. Momentum-space spin susceptibility $\chi_{\ell}(q,\omega=0)$ for (c) the $\chem{Ni}$ layer and (d) $R$ layer along $q_y=\uppi$. The dark dashed line in (d) is the single-band non-interacting Lindhard susceptibility $\chi_0$ for the bandstructure with electron density $n_0=0.125$. Measurements were obtained for $n_{\chem{Ni}}=0.875 \pm 0.001$, $n_{R}=0.125 \pm 0.001$, $U_{\chem{Ni}}=6$, $t'_{\chem{Ni}}=-0.25$, and $\beta=3$.}
\end{figure*}

\begin{figure*}[htb!]
\includegraphics[width=0.85\linewidth]{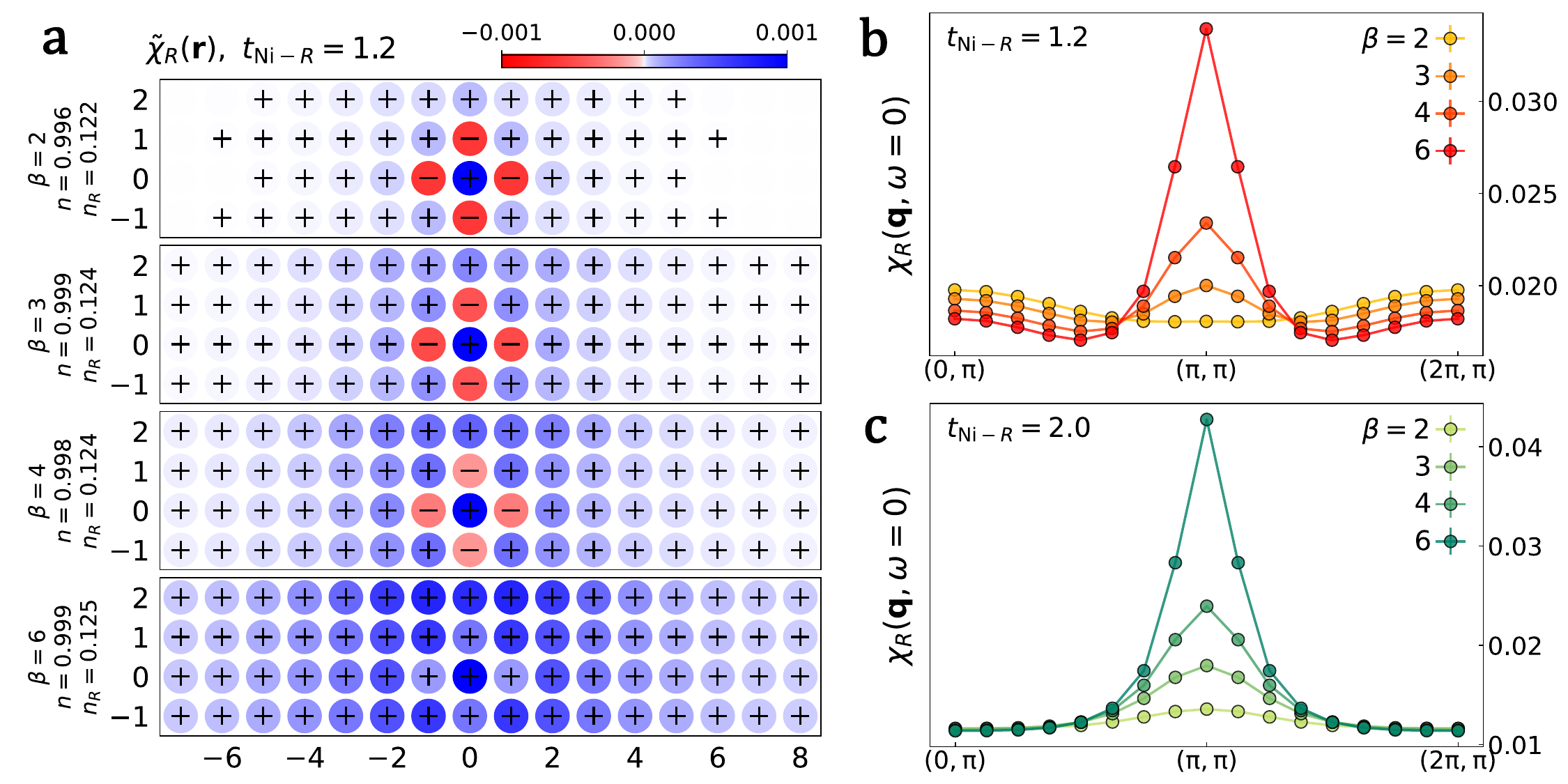}
\caption{\label{fig2} \textbf{Temperature dependence of the spin susceptibility in the $R$ layer.} (a) Real-space staggered spin susceptibility $\tilde{\chi}_{R}(\mathbf{r})$ in the $R$ layer with various inverse temperatures $\beta=1/T$ for $t_{\chem{Ni}-R}=1.2$. Momentum-space spin susceptibility in the $\chem{Ni}$ layer along $q_y=\uppi$ cut (b) for $t_{\chem{Ni}-R}=1.2$ and (c) for $t_{\chem{Ni}-R}=2$. Measurements are obtained for $n_{\chem{Ni}}=0.875$, $n_{R}=0.125$, $U_{\chem{Ni}}=6$, and $t'_{\chem{Ni}}=-0.25$.}
\end{figure*}

\section{Results}

\subsection{Spin Susceptibility}
Our desire to understand how the strong interactions of electrons are affected by the interlayer hybridization with itinerant electrons has motivated us to investigate the spin and charge density waves, which were proposed to exist in the doped single-band Hubbard model~\cite{huang2018stripe, huang2023fluctuating}.

The dynamical spin susceptibility relates to the imaginary time spin correlation functions by
\begin{equation}
\chi_{\ell}(\mathbf{q},\tau) = \int_0^{\infty} \frac{d\omega}{\uppi} \frac{\text{e}^{-\tau \omega} + \text{e}^{-(\beta-\tau)\omega}}{1-\text{e}^{-\beta \omega}} \text{Im}\chi_{\ell}(\mathbf{q},\omega),
\end{equation}
where $\chi_{\ell}(\mathbf{q},\tau)= \frac{1}{N} \sum_{i,j} \text{e}^{\mathrm{i} \mathbf{q} \cdot (\mathbf{r}_{i}-\mathbf{r}_{j})} \langle \hat{s}^{z}_{i \ell}(\tau) \hat{s}^{z}_{j \ell}(0) \rangle$. Here we only consider the $z$-component of spin $\hat{s}^{z}_{i,\ell}=\frac{1}{2}(\hat{c}_{i \uparrow}^{[\ell]\dagger}\hat{c}_{i \uparrow}^{[\ell]}-\hat{c}_{i \downarrow}^{[\ell]\dagger}\hat{c}_{i \downarrow}^{[\ell]})$ due to $SU(2)$ symmetry. In the following, we will discuss the in-plane static spin susceptibilities $\chi_{\ell}(\mathbf{q},\omega=0) = \int_0^{\beta} d\tau \chi_{\ell}(\mathbf{q},\tau)$ in the $\ell$ layer ($\ell=\chem{Ni}, R$) to study the low-energy magnetic behavior~\cite{kivelson2003detect}. Here, we note that within the range of $t_{\chem{Ni}-R}$ that we explored, the features observed in the equal-time spin correlation function are consistent with those of the static spin susceptibility.

Figure~\ref{fig1} displays the real-space staggered static spin susceptibilities within each layer at finite temperatures, {\it i.e.}, $\tilde{\chi}_{\ell}(\mathbf{r})=(-1)^{r_x+r_y}\chi_{\ell}(\mathbf{r})$, with $\mathbf{r}=(r_x,r_y)$, as well as their momentum-space counterparts, for different values of the interlayer hybridization $t_{\chem{Ni}-R}$. In Fig.~\ref{fig1}a, for small values of $t_{\chem{Ni}-R}$, spin stripes -- antiferromagnetic regions separated by anti-phase domain walls -- are found to exist in the $\chem{Ni}$ layer. It closely resembles the behavior of a doped single-band Hubbard layer, with the periodicity of the antiphase domain walls in the $\chem{Ni}$ layer matching that of the 1/8 hole-doped single-band Hubbard model, as found in Ref.~\cite{huang2018stripe, huang2023fluctuating}. In contrast, the $R$ layer closely resembles the spin susceptibility for a non-interacting single band with the same $t_{\chem{Ni}-R}$. As $t_{\chem{Ni}-R}$ increases, the AF domains of $\tilde{\chi}_{\chem{Ni}}(\mathbf{r})$ widen, and the peak of $\chi_{\chem{Ni}}(\mathbf{q})$ becomes sharper, as shown in Figure~\ref{fig1}c, indicating an increase in quasi-long-range spin correlations within the $\chem{Ni}$ layer. 

As shown in Fig.~\ref{fig1}b and \ref{fig1}d, the static spin susceptibility in the $R$ layer transitions from a non-interacting pattern to a quasi-long-range antiferromagnetic pattern for $t_{\chem{Ni}-R} > 1.4$ (see more detailed data in Supplementary Note 1). This transition shifts to smaller $t_{\chem{Ni}-R}$ as the temperature decreases. As illustrated in Fig.~\ref{fig2}, while the $R$ layer with $t_{\chem{Ni}-R}=1.2$ at temperature $T=1/2$ displays notable non-interacting features, it is dominated by uniform antiferromagnetic correlations as the temperature decreases to $T=1/6$. There is a crossover between spin stripe and uniform antiferromagnetic correlations, which curves downward towards smaller $t_{\chem{Ni}-R}$ as temperature decreases. 
Charge susceptibility measurements in Supplementary Note 1 suggest that charge-stripe behavior, which dominates the $\chem{Ni}$ layer as $t_{\chem{Ni}-R}$ is turned off, gradually disappears with increasing $t_{\chem{Ni}-R}$.


\begin{figure*}[htb!]
\includegraphics[width=1\linewidth]{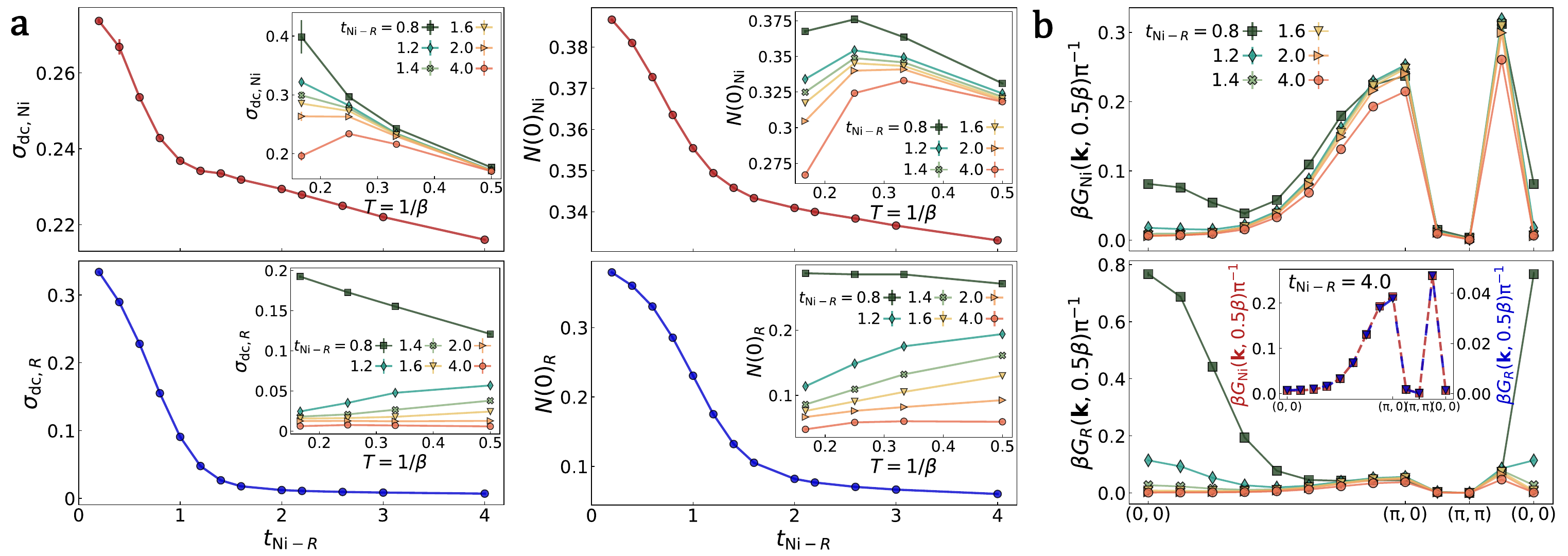}
\caption{\label{fig3} \textbf{Transport properties.} (a) DC conductivity $\sigma_{\text{dc},\ell}$ and the density of states at the Fermi energy $N(0)_{\ell}$ for the $\chem{Ni}$ layer (top panel) and $R$ layer (bottom panel) at $T=1/3$. Insets show the temperature dependence for $\beta=2, 3, 4, 6$.
(b) The proxy $\beta G_{\ell}(\mathbf{k},\tau=\frac{\beta}{2})/\uppi$ for $\ell=\chem{Ni}, R$ at the representative momentum points with $T=1/6$. The inset shows the proxies for both layers with $t_{\chem{Ni}-R}=4$ and $T=1/6$. All measurements are obtained for $n_{\chem{Ni}}=0.875 \pm 0.001$, $n_{R}=0.125 \pm 0.001$, $U_{\chem{Ni}}=6$, $t'_{\chem{Ni}}=-0.25$, and $N=16\times 4$.}
\end{figure*}

\subsection{Transport Properties}
To further investigate the mechanism by which these two layers become antiferromagnetically correlated at large $t_{\chem{Ni}-R}$, we next study the change in transport properties. To distinguish between metallic and insulating behavior, we can use the density of states through the single-particle spectral function $A(\omega)$. The imaginary time Green's function $G(\mathbf{k},\tau) = \langle \hat{c}_{\mathbf{k}}(\tau) \hat{c}^{\dagger}_{\mathbf{k}}(0) \rangle$ by
\begin{equation}
\begin{aligned}
G(\mathbf{k}, \tau) 
&= \int_{-\infty}^{\infty} d\omega \frac{\text{e}^{-\omega \tau}}{1+\text{e}^{-\beta \omega}} A(\mathbf{k}, \omega) \\
&= \int_{-\infty}^{\infty} d\omega \frac{\text{e}^{-\omega (\tau-\beta/2)}}{2 \cosh{(\beta \omega/2)}} A(\mathbf{k}, \omega),
\end{aligned}
\end{equation}
and analytic continuation can be used to determine the single-particle density of states $N(\omega) = \sum_{\mathbf{k}} A(\mathbf{k}, \omega)/N$ from the imaginary-time QMC data.
To avoid numerical complications associated with analytic continuation, we can estimate a low temperature approximation
\begin{equation}
N(0) \approx \frac{\beta}{\uppi N} \sum_{\mathbf{k}} G\left(\mathbf{k}, \tau=\frac{\beta}{2}\right).
\end{equation}

The imaginary time current-current correlator $\Lambda(\tau)$ at zero momentum can be obtained by measuring
\begin{equation}
\Lambda(\tau) = \langle \hat{\mathbf{J}}(\tau) \hat{\mathbf{J}}(0) \rangle,
\end{equation}
where $\hat{\mathbf{J}} = \mathrm{i} \sum_{i j \sigma} t_{i j} (\mathbf{r}_i - \mathbf{r}_j) \hat{c}^{\dagger}_{i \sigma} \hat{c}_{j \sigma}$.

To obtain the dc conductivity $\sigma_{\text{dc}}$ from the imaginary time response, we can either perform analytic continuation to compute $\text{Im}\Lambda(\omega)$ for all frequencies, which is related to $\Lambda(\tau)$ through the following relation:
\begin{equation}
\Lambda(\tau) = \int_{-\infty}^{\infty} \frac{d\omega}{\uppi} \frac{\text{e}^{-\omega \tau}}{1-\text{e}^{-\beta \omega}} \text{Im}\Lambda(\omega).
\end{equation}
or estimate the low frequency behavior of $\text{Im}\Lambda_{xx} \approx \omega \sigma_{\text{dc}}$ at sufficiently low temperatures from~\cite{trivedi1996superconductor}
\begin{equation}
\Lambda_{xx}\left(\tau=\frac{\beta}{2}\right) \approx \frac{\uppi \sigma_{\text{dc}}}{\beta^2}.
\end{equation}

As Fig.~\ref{fig3}a shows, the dc conductivity and the proxy for the density of states at the Fermi level for both layers manifest a decrease as $t_{\chem{Ni}-R}$ increases at $T=1/3$. As expected, both layers are metallic as long as the interlayer hybridization remains small. In the range where the $R$ layer goes from non-interacting to antiferromagnetic ($t_{\chem{Ni}-R}=1.2\sim 1.6$), the $R$ layer becomes an insulator, while the metallicity of the $\chem{Ni}$ layer exhibits a non-monotonic dependence on temperature, initially increasing with decreasing $T$ before eventually decreasing. When $t_{\chem{Ni}-R}$ is large enough, \emph{e.g.}, $t_{\chem{Ni}-R}=4$, both layers are unambiguously insulators. 

The momentum dependence of $\beta G_{\ell}(\mathbf{k}, \beta/2)/\uppi$ with $\ell=\chem{Ni}, R$ along high-symmetry cuts is shown in Fig.~\ref{fig3}b, changing with $t_{\chem{Ni}-R}$. As the proxy of the density of state at the Fermi energy, it encodes the shape of the Fermi surface. At small $t_{\chem{Ni}-R}$, the proxy suggests that the $R$ layer is likely to be a conduction band, while the $\chem{Ni}$ layer closely resembles the Hubbard band. At large $t_{\chem{Ni}-R}$, as shown in the inset, the proxies for both layers are similar in shape and differ by a constant factor.


\begin{figure}[htb!]
\includegraphics[width=1\linewidth]{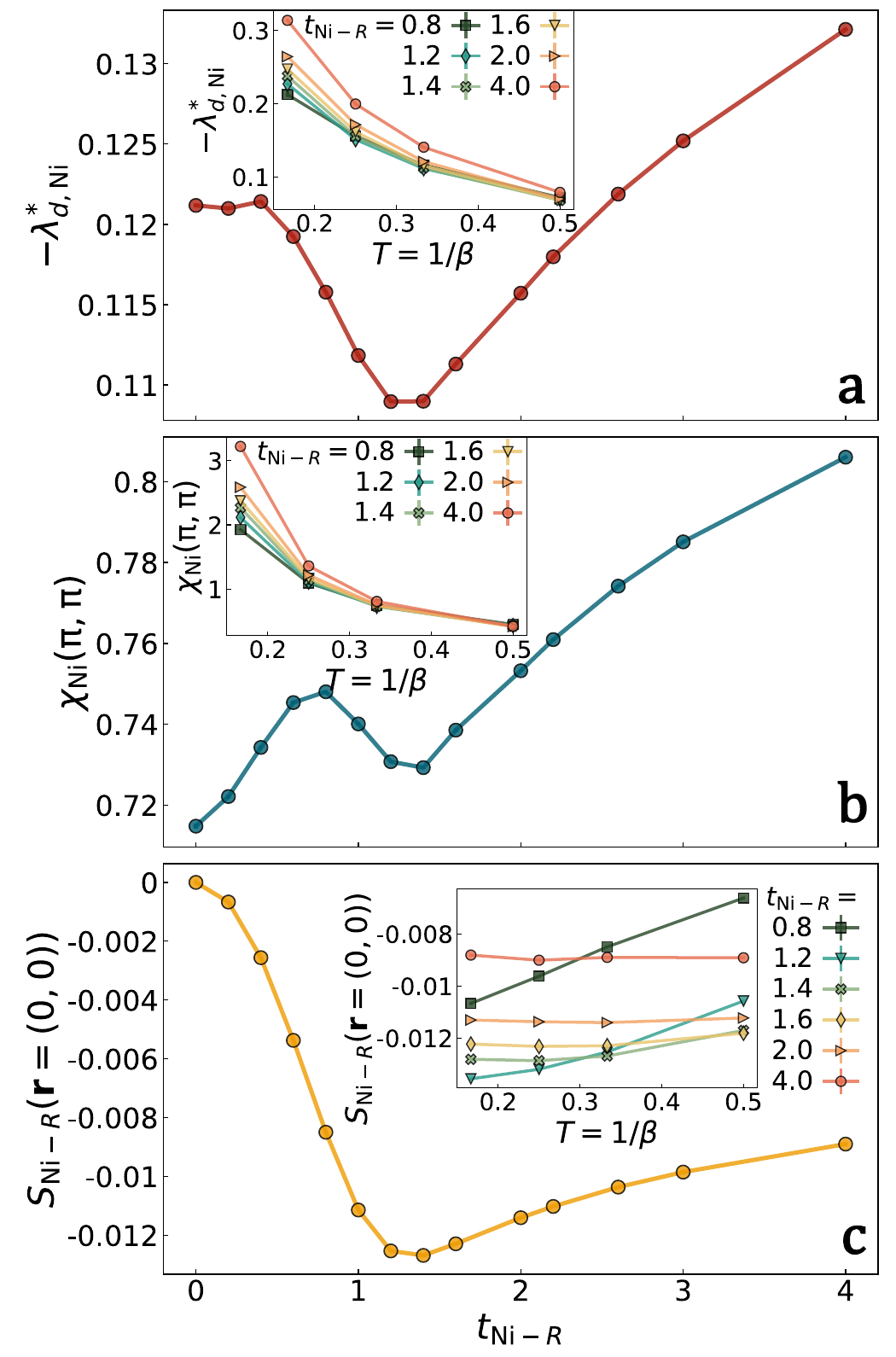}
\caption{\label{fig4} \textbf{Comparison between pair-fields, antiferromagnetic susceptibility, and on-site interlayer spin correlation.} (a) Pairing estimator with d-wave form factor in the $\chem{Ni}$ layer $\lambda^*_{d,\chem{Ni}}$, (b) static spin susceptibility at antiferromagnetic wavevector in the $\chem{Ni}$ layer $\chi_{\chem{Ni}}(\uppi,\uppi)$, and (c) equal-time on-site interlayer spin correlation, $S_{\chem{Ni}-R}(r=0)$, as functions of $t_{\chem{Ni}-R}$. Insets show temperature dependence for $1/T=2, 3, 4, 6$. All measurements are obtained for $n_{Ni}=0.875 \pm 0.001$, $n_{R}=0.125 \pm 0.001$, $U_{\chem{Ni}}=6$, $t'_{\chem{Ni}}=-0.25$, and $N=16\times 4$.}
\end{figure}

\subsection{Kondo Screening Regime} 

To explain the previous observations, we propose a heuristic understanding for the small and large interlayer hybridization regimes. 
When $t_{\chem{Ni}-R}$ is small, all electrons in the $R$ layer (accounting for $1/8$ of the electrons) are itinerant, while $7/8$ of the electrons in the $\chem{Ni}$ layer participate in forming spin stripes. 
For sufficiently strong interlayer hybridization, both a bonding- and anti-bonding-like band form, and `layer' no longer serves as a conserved index because both interacting and non-interacting bands contribute to the $\chem{Ni}$ layer and $R$ layer. 
The anti-bonding-like band is well above the Fermi energy, requiring the bonding-like band to be half-filled in order to maintain the overall electron density. This bonding-like band tends to be gapped, and consequently both layers become insulating. 
However, we will show that the system displays distinct behaviors in the intermediate  $t_{\chem{Ni}-R}$ regime, necessitating an understanding in this specific scenario.

In this section the nature of $d-$wave superconducting correlations in conjunction with antiferromagnetic correlations is explored in this model. The operator for uniform $d$-wave pairing in layer $\ell$ is commonly expressed as
\begin{equation}\label{eq:Delta_a}
\begin{aligned}
\hat{\Delta}_{d, \ell} 
&= \sum_{\mathbf{k}} (\cos{k_x}-\cos{k_y})  \hat{c}_{\mathbf{k}, \ell, \uparrow} \hat{c}_{-\mathbf{k}, \ell,\downarrow},
\end{aligned}
\end{equation}
with the pair-field susceptibility as 
\begin{equation}
P_{d, \ell} = \int_0^{\beta} d\tau \langle \hat{\Delta}_{d, \ell} (\tau) \hat{\Delta}_{d, \ell}^{\dagger} \rangle.
\end{equation}
Its disconnected part is denoted as $\bar{P}_{d, \ell}$. To reveal the dominant pairing interaction in the system, we compute the pairing vertex $\Gamma$ using the Dyson-like expression $\Gamma = P^{-1}- \bar{P}^{-1}$. We measure the pairing estimator $\lambda^*_{d,\ell} = \Gamma_{d,\ell} \bar{P}_{d,\ell}$ to represent $d$-wave pairing in layer $\ell$. 

Figure~\ref{fig4}a shows that $-\lambda^*_{d, \chem{Ni}}$ has a valley minimum around $t_{\chem{Ni}-R}\sim 1.2$ and rises monotonically as $t_{\chem{Ni}-R}$ increases for $T=1/3$. Small $t_{\chem{Ni}-R}$ perturbatively generates a Kondo-like coupling, so Kondo singlets may form in the regime where Kondo coupling becomes competitive with intralayer antiferromagnetic Heisenberg coupling, suppressing $d$-wave pairing and uniform in-plane antiferromagnetic correlations. The appearance of a similar valley in the antiferromagnetic spin susceptibility $\chi_{\chem{Ni}}(\uppi,\uppi)$, as shown in Fig.~\ref{fig4}b, and the equal-time on-site interlayer spin correlation $S_{\chem{Ni}-R}\left(\mathbf{r}=\left(0,0\right)\right) = \chi_{\chem{Ni}-R}\left(\mathbf{r}=\left(0,0\right),\tau=0\right)$, which is an indication of the formation of interlayer singlets, further support this idea. As shown in Fig.~\ref{fig4}c, interlayer singlet formation tends to peak at $t_{\chem{Ni}-R} \sim 1.2$, coinciding with the suppression of $d$-wave pairing and antiferromagnetic correlations. 
Notice that the $t_{\chem{Ni}-R}$ of the valley shifts towards smaller values as the temperature decreases. Additional details can be found in Supplementary Notes 2-4.

Our observations suggest that, at intermediate $t_{\chem{Ni}-R}$, $1/8$ of the electrons in the $\chem{Ni}$ layer are screened by itinerant electrons from the $R$ layer, forming Kondo singlets, leaving $6/8$ of the electrons in the $\chem{Ni}$ layer as charge carriers. Meanwhile, the $R$ layer, being devoid of charge carriers, exhibits a significant reduction in the density of states at the Fermi energy, which is consistent with our observations in Fig.~\ref{fig3}. 

\section{Discussion}
A clear cross-over from spin and charge stripe-like behavior to uniform antiferromagnetic correlations in the interacting $\chem{Ni}$ layer occurs as one increases the interlayer hybridization $t_{\chem{Ni}-R}$ in the model. Even the non-interacting $R$ layer develops uniform antiferromagnetic correlations as the interlayer hybridization becomes large enough for a given set of parameters. Focusing on the region near the cross-over, the antiferromagnetic spin susceptibility, as well as the proxy of uniform $d$-wave pair-field susceptibility ($-\lambda^*_{d, \chem{Ni}}$), exhibit a suppression in the $\chem{Ni}$ layer, where an effective (perturbative) interlayer Kondo coupling $J_{\text{K}}\sim 2t_{\chem{Ni}-R}^2/U_{\chem{Ni}}$ becomes competitive with an effective (perturbative) intralayer Heisenberg coupling $J_{\text{H}}\sim 4t_{\chem{Ni}}^2/U_{\chem{Ni}}$. 
Such a scenario may lead to screening of the electrons in the $\chem{Ni}$ layer by all of the available itinerant electrons in the $R$ layer, resulting in the formation of Kondo singlets and quenching the itinerant charge degrees of freedom. Our observations of the dramatic decrease in conductivity in the $R$ layer, the suppression in intralayer antiferromagnetic susceptibility and d-wave pair fields induced by magnetic fluctuations, and more importantly, an enhancement of the on-site interlayer spin-spin correlation, support this hypothesis. 
For sufficiently strong $t_{\chem{Ni}-R}$, the Hubbard interaction projected into the half-filled bonding-orbital band induces strong antiferromagnetic correlations and insulating behavior.

\section{Acknowledgements}
This work was supported by the U.S. Department of Energy (DOE), Office of Basic Energy Sciences, Division of Materials Sciences and Engineering under Contract No. DE-AC02-76SF00515. EWH was supported by the Gordon and Betty Moore Foundation’s EPiQS Initiative through grants GBMF 4305 and GBMF 8691 at the University of Illinois Urbana-Champaign. CJ acknowledges the support from U.S. Department of Energy, Office of Science, Basic Energy Sciences under Award No. DE-SC0022216. Computational work was performed on the Sherlock cluster at Stanford University and on resources of the National Energy Research Scientific Computing Center, supported by the U.S. DOE, Office of Science, under Contract no. DE-AC02-05CH11231.

\bibliography{ref.bib}

\end{document}


\title{\Large Supplementary Information\\
\large The emergence of antiferromagnetic correlations and Kondo-like features\\
in a two-band model for infinite-layer nickelates}

\author{Fangze Liu}
\affiliation{Department of Physics, Stanford University, Stanford, California 94305, USA}
\affiliation{Stanford Institute for Materials and Energy Sciences, SLAC National Accelerator Laboratory, 2575 Sand Hill Road, Menlo Park, CA 94025, USA}

\author{Cheng Peng}
\affiliation{Stanford Institute for Materials and Energy Sciences, SLAC National Accelerator Laboratory, 2575 Sand Hill Road, Menlo Park, CA 94025, USA}

\author{Edwin W. Huang}
\affiliation{Department of Physics and Institute of Condensed Matter Theory, University of Illinois at Urbana-Champaign, Urbana, Illinois 61801, USA}
\affiliation{Department of Physics and Astronomy, University of Notre Dame, Notre Dame, IN 46556, United States}
\affiliation{Stavropoulos Center for Complex Quantum Matter, University of Notre Dame, Notre Dame, IN 46556, United States}

\author{Brian Moritz}
\affiliation{Stanford Institute for Materials and Energy Sciences, SLAC National Accelerator Laboratory, 2575 Sand Hill Road, Menlo Park, CA 94025, USA}

\author{Chunjing Jia}
\affiliation{Stanford Institute for Materials and Energy Sciences, SLAC National Accelerator Laboratory, 2575 Sand Hill Road, Menlo Park, CA 94025, USA}

\author{Thomas P. Devereaux}
\affiliation{Stanford Institute for Materials and Energy Sciences, SLAC National Accelerator Laboratory, 2575 Sand Hill Road, Menlo Park, CA 94025, USA}
\affiliation{Department of Materials Science and Engineering, Stanford University, CA 94305, USA}

\maketitle

\renewcommand{\thesection}{\arabic{section}}
\renewcommand{\thefigure}{\arabic{figure}}

In this Supplemental Material, we present additional data and analyses. 
Supplementary Note \ref{supp1} provides extensive measurements of intralayer static spin susceptibilities, Lindhard susceptibilities, and intralayer charge susceptibilities, which supplement the figures in the main text.
Supplementary Note \ref{supp2} shows a detailed plot demonstrating the temperature dependence of the intralayer pairing vertex and long-range antiferromagnetic susceptibility, and interlayer spin responses in relation to interlayer hopping.
Supplementary Note \ref{supp3} discusses the distinction between on-site interlayer static spin susceptibility and equal-time spin correlation.

\section{Static spin and charge susceptibilities} \label{supp1}
\label{supp1}

In the main text, we mainly discuss the static spin susceptibilities for representative values of $t_{\chem{Ni}-R}$. Supplementary Figure \ref{sfig1} shows the spin susceptibilities along $q_y = 0$ for different values of $t_{\chem{Ni}-R}$.
In Supplementary Figure \ref{sfig2} we show a more comprehensive $t_{\chem{Ni}-R}$ dependence of intralayer spin susceptibilities in both layers. We also present the staggered Lindhard susceptibility, denoted as $\chi_{\ell}^0(\bold{r})$, for our two-band Hamiltonian without interaction, utilizing the same band parameters, including $t_{\chem{Ni}-R}$, and for the same electron densities.

\begin{figure}[H]
\includegraphics[width=0.79\linewidth]{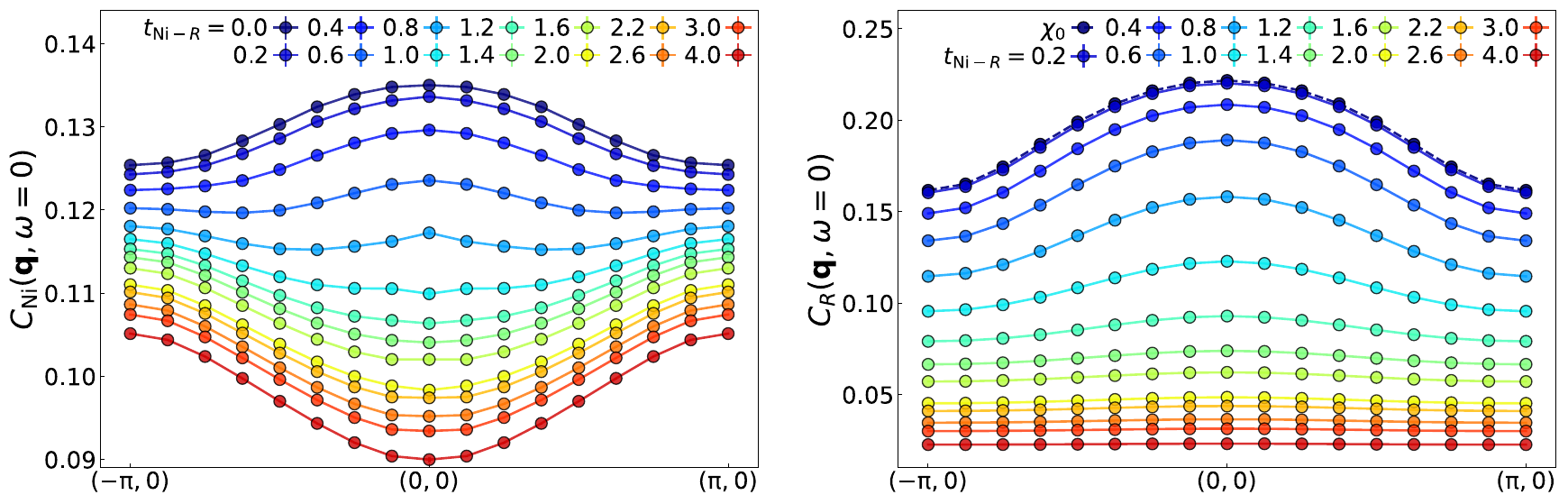}
\caption{\label{sfig3} Momentum-space charge susceptibility $C_{\ell}(\bold{q},\omega=0)$ for the $\chem{Ni}$ layer (left) and $R$ layer (right) for various $t_{\chem{Ni}-R}$ along $q_y=0$. The dark dashed line $\chi_0$ is the single-band non-interacting Lindhard charge susceptibility with electron density $n_0=0.125$. Measurements are obtained with $n_{\chem{Ni}}=0.875 \pm 0.001$, $n_{R}=0.125 \pm 0.001$, $U_{\chem{Ni}}=6$, $t'_{\chem{Ni}}=-0.25$, and $\beta=3$.} 
\end{figure}

\begin{figure}[H]
\includegraphics[width=0.79\linewidth]{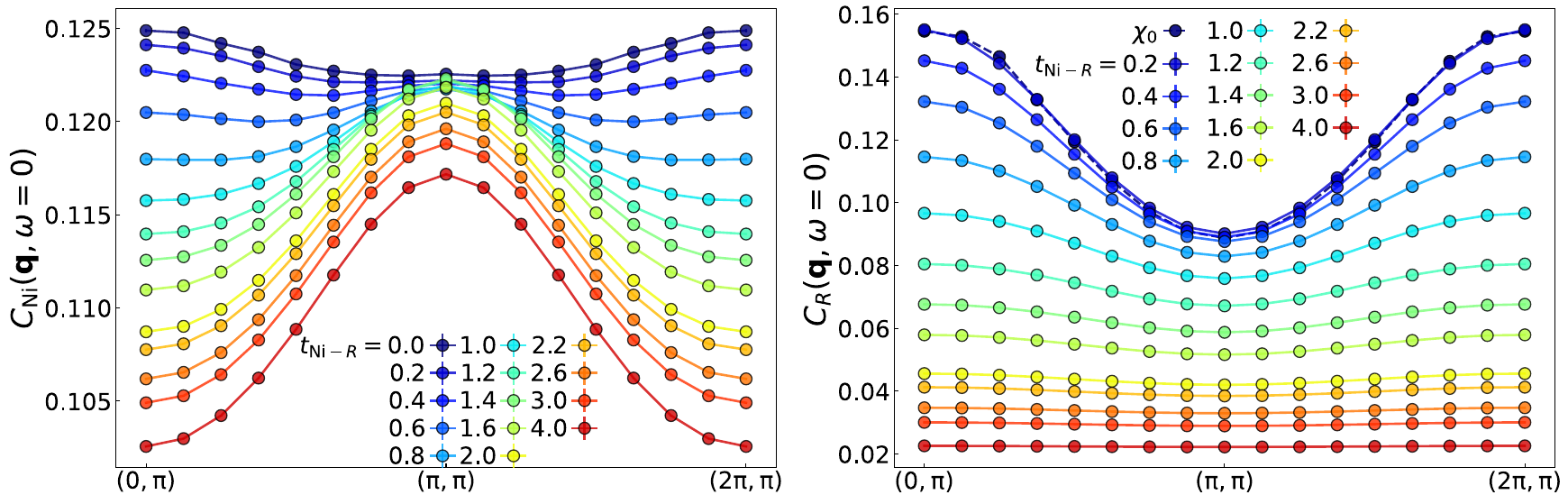}
\caption{\label{sfig4} Momentum-space charge susceptibility $C_{\ell}(\bold{q},\omega=0)$ for the $\chem{Ni}$ layer (left) and $R$ layer (right) for various $t_{\chem{Ni}-R}$ along $q_y=\uppi$. The dark dashed line $\chi_0$ is the single-band non-interacting Lindhard charge susceptibility with electron density $n_0=0.125$. Measurements are obtained with $n_{\chem{Ni}}=0.875 \pm 0.001$, $n_{R}=0.125 \pm 0.001$, $U_{\chem{Ni}}=6$, $t'_{\chem{Ni}}=-0.25$, and $\beta=3$.}
\end{figure}

\begin{figure}[H]
\includegraphics[width=0.55\linewidth]{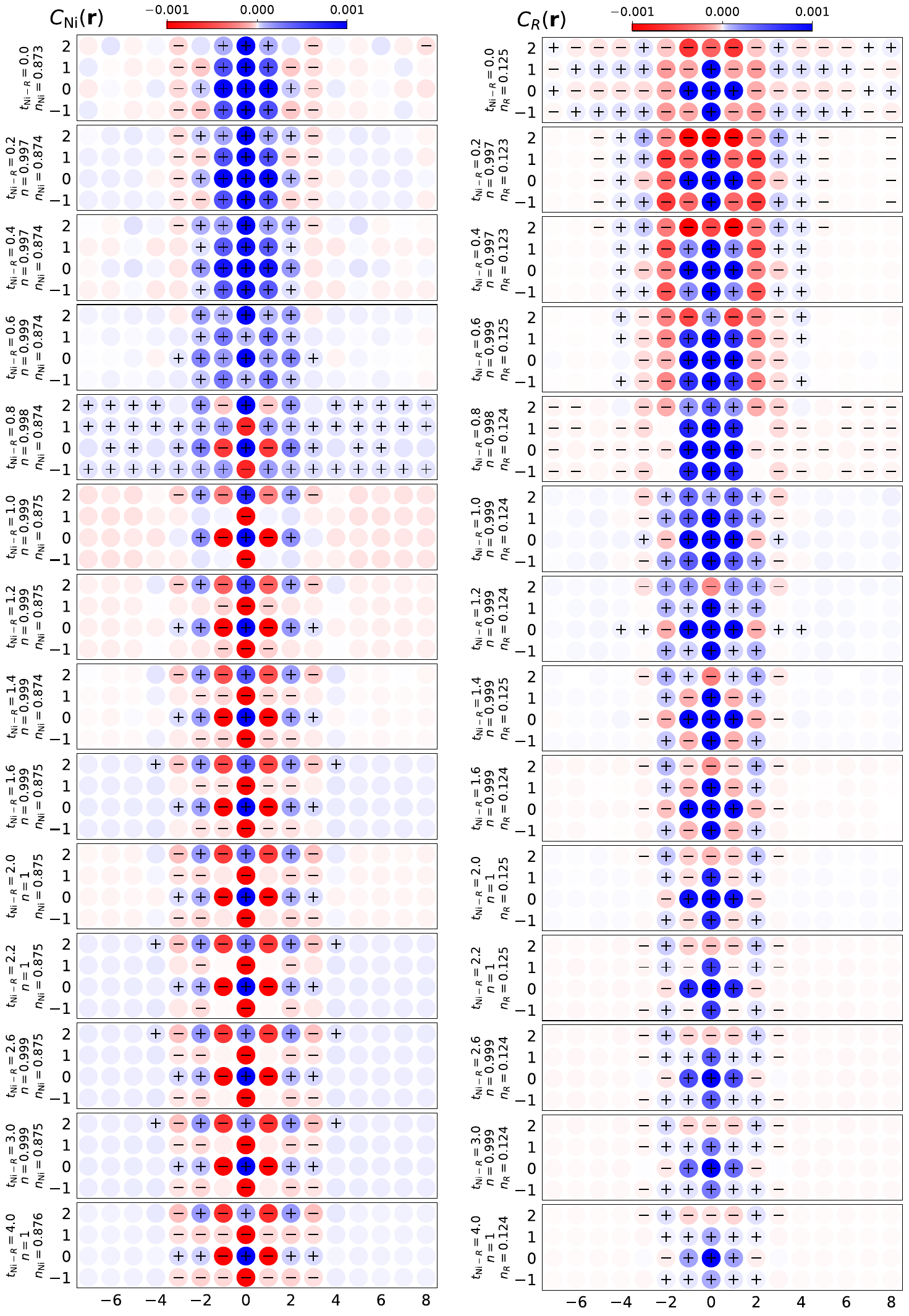}
\caption{\label{sfig5} Real-space charge susceptibilities in the $\chem{Ni}$ and $R$ layer for various $t_{\chem{Ni}-R}$. A $+$ or $-$ are shown when the susceptibilities are nonzero by at least two standard errors. Measurements are obtained with $n_{\chem{Ni}}=0.875 \pm 0.001$, $n_{R}=0.125 \pm 0.001$, $U_{\chem{Ni}}=6$, $t'_{\chem{Ni}}=-0.25$, and $\beta=3$.}
\end{figure}

Here, we also study the charge density waves in the two-band model by calculating the static charge susceptibility in the $\ell$ layer ($\ell=\{\chem{Ni}, R\}$):
\begin{equation}
\begin{aligned}
& C_{\ell}(\bold{q}, \omega=0) = \int_{0}^{\beta} d\tau C_{\ell}(\bold{q}, \tau),\\
& C_{\ell}(\bold{q}, \tau) = \frac{1}{N} \sum_{i,j} \text{e}^{\mathrm{i} \bold{q} \cdot (\bold{r}_{i}-\bold{r}_{j})} \Big(\langle \hat{n}_{i}^{[\ell]}(\tau) \hat{n}_{j}^{[\ell]}(0) \rangle - \langle \hat{n}_{i}^{[\ell]}(\tau) \rangle \langle \hat{n}_{j}^{[\ell]}(0) \rangle \Big),
\end{aligned}
\end{equation}
where $\hat{n}_{i}^{[\ell]} = \sum_{\sigma} \hat{c}^{[\ell] \dagger}_{i, \sigma} \hat{c}^{[\ell]}_{i, \sigma}$. We plot the momentum-space (Supplementary Figure \ref{sfig3} and Supplementary Figure \ref{sfig4}) and real-space (Supplementary Figure \ref{sfig5}) static charge susceptibilities for various $t_{\chem{Ni}-R}$.

\begin{figure}[H]
\includegraphics[width=0.79\linewidth]{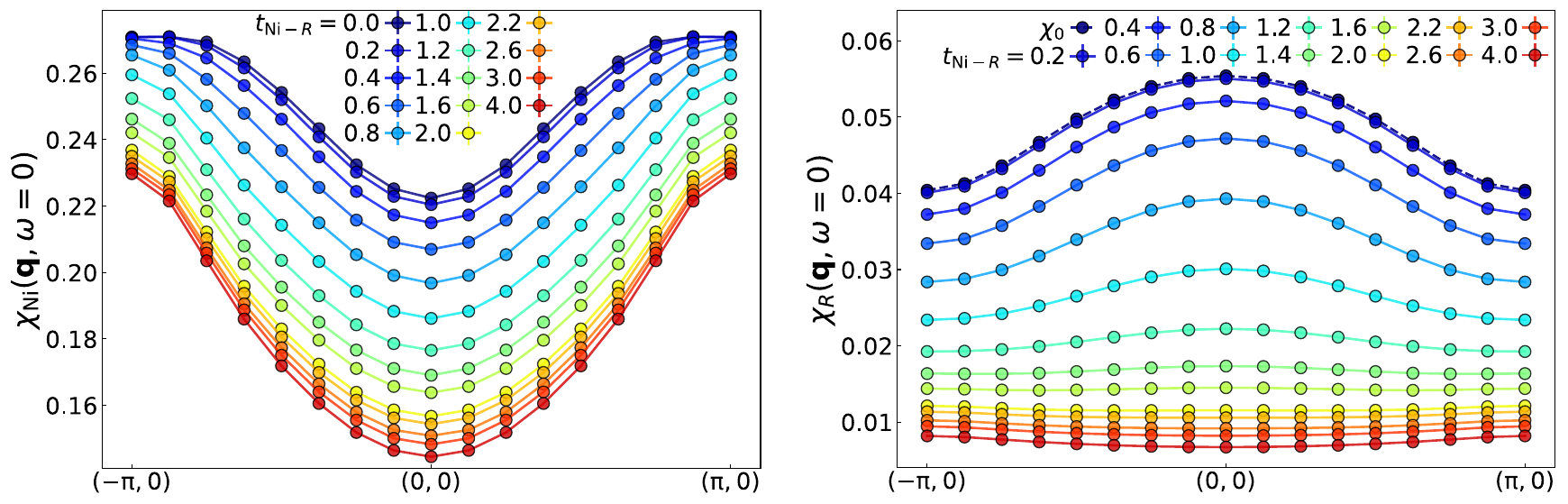}
\caption{\label{sfig1} Momentum-space spin susceptibility $\chi_{\ell}(\bold{q},\omega=0)$ for the $\chem{Ni}$ layer (left) and $R$ layer (right) for various $t_{\chem{Ni}-R}$ along $q_y=0$. The dark dashed line $\chi_0$ is the single-band non-interacting Lindhard charge susceptibility with electron density $n_0=0.125$. Measurements are obtained with $n_{\chem{Ni}}=0.875 \pm 0.001$, $n_{R}=0.125 \pm 0.001$, $U_{\chem{Ni}}=6$, $t'_{\chem{Ni}}=-0.25$, and $\beta=3$.}
\end{figure}

\begin{figure}[H]
\includegraphics[width=1\linewidth]{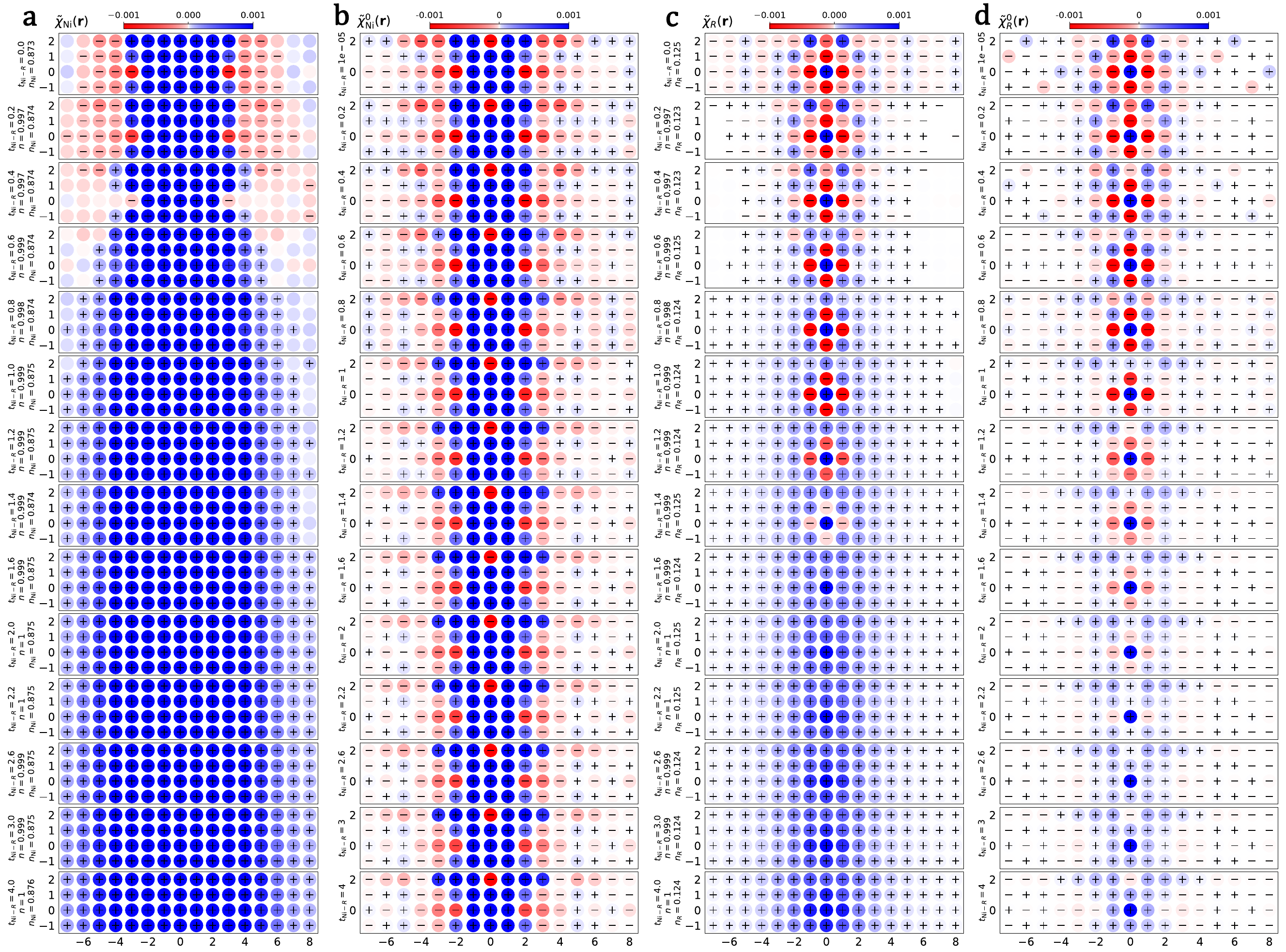}
\caption{\label{sfig2} Real-space staggered spin susceptibilities (a) $\tilde{\chi}_{\text{Ni}}(\bold{r})$ in the $\chem{Ni}$ band and (c) $\tilde{\chi}_{\text{R}}(\bold{r})$ in the $R$ band, and staggered Lindhard susceptibilities (b) $\tilde{\chi}_{\text{Ni}}^0(\bold{r})$ in the $\chem{Ni}$ band and (d) $\tilde{\chi}_{\text{R}}^0(\bold{r})$ in the $R$ band, for various $t_{\chem{Ni}-R}$.
A $+$ or $-$ are shown when the susceptibilities are nonzero by at least two standard errors. Measurements are obtained with $n_{\chem{Ni}}=0.875 \pm 0.001$, $n_{R}=0.125 \pm 0.001$, $U_{\chem{Ni}}=6$, $t'_{\chem{Ni}}=-0.25$, and $\beta=3$.}
\end{figure}

\section{Temperature dependence of pairing and spin responses}\label{supp2}

\begin{figure}[H]
\includegraphics[width=0.8\linewidth]{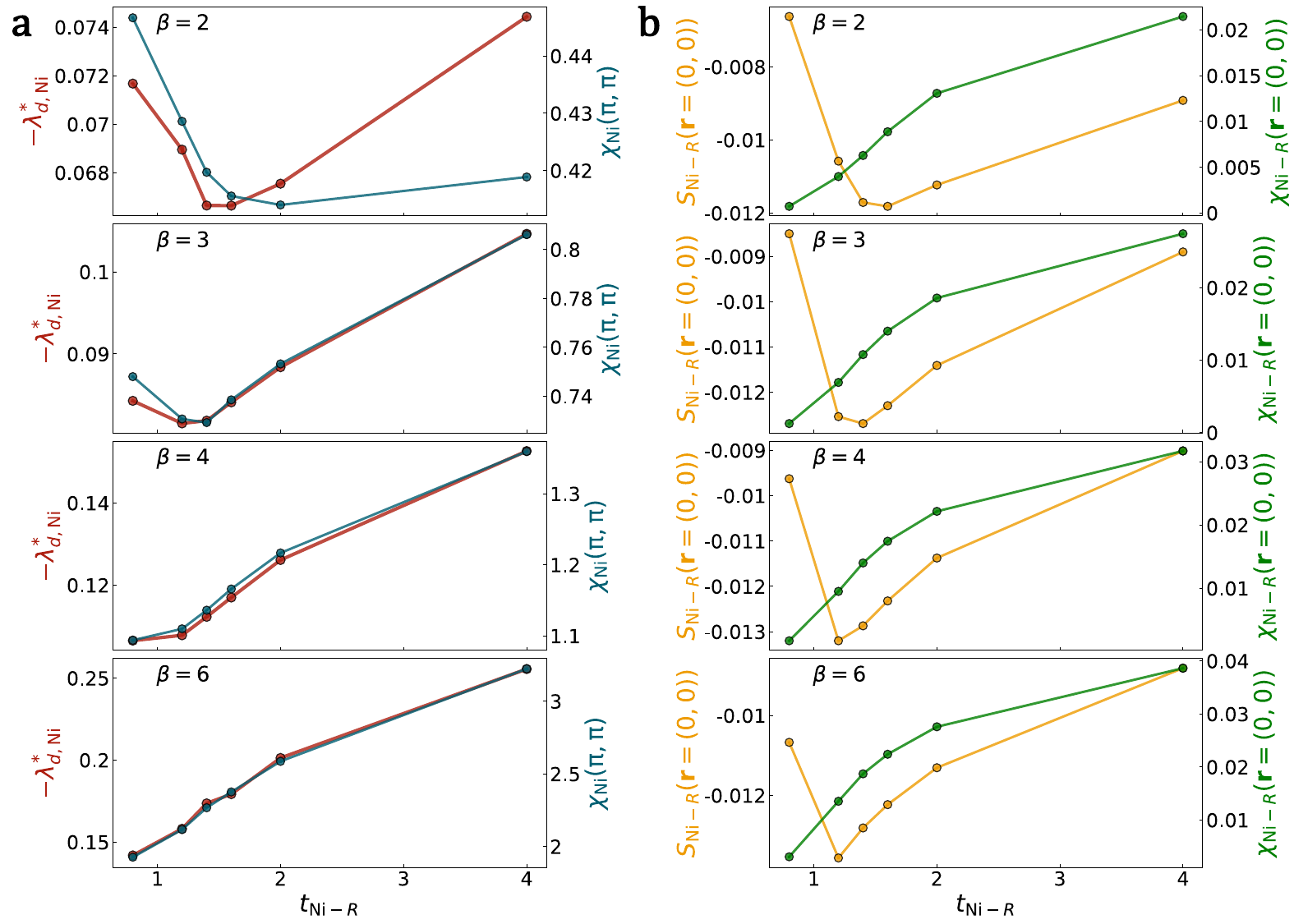}
\caption{\label{sfig6} $t_{\chem{Ni}-R}$ dependence of (a) uniform $d$-wave pairing estimator $-\lambda^*_{d,\chem{Ni}}$ and static spin susceptibilities in the $\chem{Ni}$ layer at $\bold{q}=(\uppi,\uppi)$, and (b) on-site interlayer $\chi_{\chem{Ni}-R}\left(\bold{r}=\left(0,0\right),\omega=0\right)$ with $\beta=2,3,4,6$. All measurements are obtained with $n_{\chem{Ni}}=0.875 \pm 0.001$, $n_{R}=0.125 \pm 0.001$, $U_{\chem{Ni}}=6$, and $t'_{\chem{Ni}}=-0.25$.}
\end{figure}

\section{On-site interlayer spin response}\label{supp3}

To further demonstrate the relationship between singlet formation, pairing, and antiferromagnetic correlations at various temperatures, we present a plot of these observables as a function of $t_{\chem{Ni}-R}$ at different temperatures in Supplementary Figure \ref{sfig6}. We further include the on-site interlayer spin susceptibility $\chi_{\chem{Ni}-R}\left(\bold{r}=\left(0,0\right)\right)$, which exhibits a distinct behavior and opposite sign compared to $S_{\chem{Ni}-R}\left(\bold{r}=\left(0,0\right)\right)$. Note that the intralayer antiferromagnetic spin susceptibility exhibit the same behavior as the spin-spin correlation within the scope of our study.

To better understand this difference, we perform exact diagonalization (ED) on an one-site bilayer Hubbard model, using the same Hamiltonian as in the main text. Considering that DQMC is formulated in the grand canonical ensemble, we do not conserve the total number of electrons. We keep the model parameters constant ($U_{\chem{Ni}}=6$, $t_{\chem{Ni}-R}=4$) and adjust the chemical potentials to obtain $1/8$-hole doping in the $\chem{Ni}$ layer and $1/8$-electron filling in the $R$ layer for the ground state. The following table presents the eigenstates with the lowest eigenvalues and their corresponding physical observables:

\begin{center}\label{table:ED_h0}
\begin{tabular}{ c | c | c | c | c | c | c }
 Eigenvalue & Eigenvector & $\langle \hat{S}_{\chem{Ni}}^z \hat{S}_{R}^z \rangle$ & $\langle \hat{S}_{R}^z \rangle$ & $\langle \hat{n}_{\chem{Ni}} \rangle$ &  $\langle \hat{n}_{R} \rangle$ & $\langle \hat{n} \rangle$\\
 \hline
$-1.508$ & $ 0.303 \mtrxud{0}{\downarrow} + 0.114 \mtrxud{\downarrow}{0} +0.885 \mtrxud{0}{\uparrow} + 0.334\mtrxud{\uparrow}{0} $ & $0$ & $0.049$ & $0.875$ & $0.125$ & $1$ \\
 \hline
$-1.508$ & $ 0.885 \mtrxud{0}{\downarrow} + 0.334 \mtrxud{\downarrow}{0} - 0.303 \mtrxud{0}{\uparrow} - 0.114\mtrxud{\uparrow}{0} $ & $0$ & $-0.049$ & $0.875$ & $0.125$ & $1$ \\
 \hline
 $0$ & $ 1 \mtrxud{0}{0}$ & $0$ & $0$ & $0$ & $0$ & $0$ \\
\hline
 $0.932$ & $ 0.728 \mtrxud{0}{\uparrow \downarrow} +0.461 \mtrxud{\downarrow}{\uparrow} +0.461 \mtrxud{\uparrow}{\downarrow} + 0.214\mtrxud{\uparrow\downarrow}{0} $ & $-0.106$ & $0$ & $1.485$ & $0.517$ & $2$ \\
 \hline
.... &  ... &  ... &  ... &  ... &  ... &  ... \\
\end{tabular}
\end{center}

In the table, the basis vector, for example, $\mtrxud{\uparrow \downarrow}{\downarrow}$ represents that there are two electrons in the $R$ layer, one with spin up and one with spin down, and one electron with spin down in the $\chem{Ni}$ layer. \\

We observe a two-fold degenerate ground state at the $n=1$ sector. Focusing solely on the ground state, both $\langle \hat{S}_{\chem{Ni}}^z \hat{S}_{R}^z \rangle$ and $\langle \hat{S}_{R}^z \rangle$ are zero. The fourth row of the table demonstrates that the eigenstates in the $n=2$ sector contribute to a negative value to $\langle \hat{S}_{\chem{Ni}}^z \hat{S}_{R}^z \rangle$. Consequently, the ensemble average of $S_{\chem{Ni}}^z S_{R}^z$ is negative, while that of $S_{R}^z$ remains zero. 

Herein, we introduce a perturbative magnetic field $h_{\chem{Ni}}$ to the $\chem{Ni}$ layer. We then observe the resulting spin magnetization in the $R$ layer, $\langle \hat{S}_{R}^z \rangle$, and then the spin susceptibility between the $\chem{Ni}$ layer and the $R$ layer can be calculated by $\chi_{\chem{Ni}-R}=\frac{\partial \langle \hat{S}_{R}^{z} \rangle}{\partial h_{\chem{Ni}}}$ (it is equivalent to $\chi_{\chem{Ni}-R}(\omega=0) = \int_0^{\beta} d\tau \langle \hat{S}^{z}_{\chem{Ni}}(\tau) \hat{S}^{z}_{R}(0) \rangle $ as defined in the main text based on fluctuation–dissipation theorem). The following table displays the eigenstates and their corresponding physical observables with $h_{\chem{Ni}}=0.1$:

\begin{center}\label{table:ED_h0.1}
\begin{tabular}{ c | c | c | c | c | c | c }
 Eigenvalue & Eigenvector & $\langle \hat{S}_{\chem{Ni}}^z \hat{S}_{R}^z \rangle$ & $\langle \hat{S}_{R}^z \rangle$ & $\langle \hat{n}_{\chem{Ni}} \rangle$ &  $\langle \hat{n}_{R} \rangle$ & $\langle \hat{n} \rangle$\\
 \hline
$-1.552$ & $ 0.936 \mtrxud{0}{\uparrow} + 0.352\mtrxud{\uparrow}{0} $ & $0$ & $0.062$ & $0.876$ & $0.124$ & $1$ \\
 \hline
$-1.502$ & $ 0.936 \mtrxud{0}{\downarrow} + 0.352 \mtrxud{\downarrow}{0} $ & $0$ & $-0.062$ & $0.875$ & $0.125$ & $1$ \\
 \hline 
 $0$ & $ 1 \mtrxud{0}{0}$ & $0$ & $0$ & $0$ & $0$ & $0$ \\
\hline
 $0.932$ & $ 0.728 \mtrxud{0}{\uparrow \downarrow} +0.464 \mtrxud{\downarrow}{\uparrow} +0.458 \mtrxud{\uparrow}{\downarrow} + 0.214\mtrxud{\uparrow\downarrow}{0} $ & $-0.106$ & $-0.003$ & $1.485$ & $0.517$ & $2$ \\
 \hline
.... &  ... &  ... &  ... &  ... &  ... &  ... \\
\end{tabular}
\end{center}

We find that the magnetic field lifts the degeneracy of the ground state, so it leads to the ground state with a positive $\langle \hat{S}_{R}^z \rangle$ having the largest Boltzmann weight, which results in a positive ensemble average of $S_{R}^z$. Therefore, we conclude that magnetic field favors a spin-up electron in the $\chem{Ni}$ layer. Focusing on the $n=1$ sector, the eigenstate with $\ket{\psi_0} = 0.936 \mtrxud{\uparrow}{0} + 0.352 \mtrxud{0}{\uparrow}$ emerges as having the lowest eigenvalue, which leads to a positive value of on-site interlayer spin susceptibility $\chi_{\chem{Ni}-R}\left(\bold{r}=\left(0,0\right)\right)$. The other notable observation is that $\chi_{\chem{Ni}-R}\left(\bold{r}=\left(0,0\right)\right)$ calculated through Lehmann representation in the system without perturbative applied field is also positive at all finite temperatures. \\

In conclusion, the single-site ED analysis offers a complementary explanation for the distinct behaviors of the on-site interlayer spin correlation $S_{\chem{Ni}-R}\left(\bold{r}=\left(0,0\right)\right)$ and spin susceptibility $\chi_{\chem{Ni}-R}\left(\bold{r}=\left(0,0\right)\right)$. Specifically, the eigenstates with two electrons per site collectively yield a negative value of $S_{\chem{Ni}-R}\left(\bold{r}=\left(0,0\right)\right)$, while the eigenstates with one electron per site result in a positive value of $\chi_{\chem{Ni}-R}\left(\bold{r}=\left(0,0\right)\right)$. Essentially, this difference arises from the fact that the equal-time correlation is more affected by high-energy excitations than the static susceptibility.